# User Recommendation System based on MIND Dataset

**Niran A. Abdulhussein, Ahmed J. Obaid**
nirana.alasadi@student.uokufa.edu.iq ; ahmedj.aljanaby@uokufa.edu.iq
**Faculty of Computer Science and Mathematics, University of Kufa, Iraq**

**Abstract**
Nowadays, it's a very significant way for researchers and other individuals to achieve their interests because it provides short solutions to satisfy their demands. Because there are so many pieces of information on the internet, news recommendation systems allow us to filter content and deliver it to the user in proportion to his desires and interests. RSs have three techniques: content-based filtering, collaborative filtering, and hybrid filtering. We will use the MIND dataset with our system, which was collected in 2019, the big challenge in this dataset because there is a lot of ambiguity and complex text processing. In this paper, will present our proposed recommendation system. The core of our system we have used the GloVe algorithm for word embeddings and representation. Besides, the Multi-head Attention Layer calculates the attention of words, to generate a list of recommended news. Finally, we achieve good results more than some other related works in AUC 71.211, MRR 35.72, nDCG@5 38.05, and nDCG@10 44.45.

## 1. Introduction

In this study, we will present our proposed system for RS, which we suggested and apply to our dataset (MIND dataset) which was an enormous number of texts that were difficult to process. This system consists of **5** phases: The first phase is Data Collection and splitting. The second phase includes preprocessing of our data. The third phase included the word embedding and representation that depends on a GloVe algorithm. The fourth phase is finding the candidate news from the MIND dataset. And the last phase we can find the list of recommended news for users by comparing the candidate list with the recent news.

After this, we will discuss the results that have been produced from our implementation of the proposed system and methodology, and System Evaluation and Results Discussion. So, we applied this model to the MIND dataset collected from the Microsoft news platform for English News articles. Choosing a set of steps and algorithms that have a direct impact on determining the accuracy of our system.

We collected our data from the Microsoft News platform site for **6** weeks, the **MI**crosoft **N**ews **D**ataset (**MIND**[1]) is a large-scale dataset for news recommendation research. It was developed using anonymized Microsoft News website behavior records. The goal of MIND is to serve as a benchmark dataset for news recommendation and to support research in the field of news recommendation and recommender systems.

MIND now has around **160k** English news articles and over **15** million impression logs produced by **one** million users. Every news item contains extensive textual content such as a title, abstract, body, category, and entities. Each impression log includes historical click activities, non-clicked activities, and history news click behavior for this user. Each user was securely hashed into an anonymized ID and de-linked from the production system to preserve user privacy [1]. MIND are metrics in **AUC**[2], **MRR**[3], **nDCG@5**[4], and **nDCG@10** as the evaluation metrics.

---
[1] https://msnews.github.io/

Table 1: MIND dataset statistics analysis.

| Dataset Statistics | | | |
|---|---|---|---|
| No. of Users | 1,000,000 | No. of Topic categories | 20 |
| No. of News | 161,013 | Avg. of No. of words | 11-12 |
| No. of Impressions | 15,777,377 | Avg. of No. of entities | 16-17 |
| No. of Entities | 3,299,687 | No. of Versions | 2 |
| No. of Click behaviors | 24,155,470 | No. of Labels | 8 |
| **Dataset Characteristics** | **Training set** | **Testing set** | **Total** |
| No. of Instances for News | 101,527 records | 120,959 records | 222,486 records |
| No. of Instances for Behaviors | 1,000,000 records | 1,048,576 records | 2,048,576 records |
| No. of Definite article | 345,712 | 412,704 | 758,416 |
| No. of Auxiliary verbs | 196,547 | 234,325 | 430,872 |
| No. of pronouns | 237,340 | 283,137 | 520,477 |
| No. of prepositions | 720,058 | 860,446 | 1,580,504 |
| No. of words | 14,090,156 | 12,177,718 | 26,267,874 |
| Size | 530.2 MB | 605 MB | 1135.2 MB |

## 2. Related Works

In this section, we will present the most related works of the researchers in previous studies about recommendation system research that are based on the MIND-large dataset on Microsoft News platforms, with the increase of the data on the Internet, recommendation algorithms that are used with this dataset, and the performance of models that used with the year of study in each experiment's results.

The most related works researchers that work on this dataset on the MIND leaderboard in state of the art as the first team are the Microsoft teams represented by Chuhan Wu and et al. in (2021) used an ensemble of additive attention (Fastformer+PLM-NR) model achieved the best AUC 72.68, MRR 37.45, nDCG@5 46.84, and nDCG@10 41.51 [2]. Yu Song and et al. in (2021) used a progressive hierarchical user Contextual bandit (pHUCB) model that achieved AUC 0.723 [3]. Jian Li and et al. in (2022) used Multi-Interest Matching Network for News Recommendation (MINER) which achieved AUC 71.51, MRR 36.18, nDCG@5 39.72, nDCG@10 45.34 [4]. U Kang in (2020) used an ensembler consisting of two NNG + four NNB layered and achieved AUC

---
[2] Area Under the ROC Curve.
[3] Monthly Recurring Revenue.
[4] The normalized participant's DCG score (Discounted Cumulative Gain) divided by the ideal ranking's DCG score yields the nDCG score.

0.7114, MRR 0.3568, nDCG@5 0.3916, nDCG@10 0.4485 [5]. Chuhan Wu and et al. in (2021) used the Federated learning method based on Knowledge Distillation (FedKD) and achieved AUC 71.0, MRR 35.6, nDCG@5 38.9, nDCG@10 44.8 [6]. Qi Zhang and et al. in (2021) used User-News BERT (UNBERT) achieved AUC 0.7068, MRR 0.3568, nDCG@5 0.3913, nDCG@10 0.4478 [7]. Chuhan Wu and et al. in (2021) used pre-trained language models (PLMs) and achieved AUC 70.64, MRR 35.39, nDCG@5 38.71, nDCG@10 44.38 [8]. Shuqi Lu and et al. in (2022) used Strong tExt Encoder by training with weak Decoder (SEED-Encoder) and achieved AUC 0.7059, MRR 0.3506, nDCG@5 0.3908, nDCG@10 0.4526 [9]. Chuhan Wu and et al. in (2021) used News-BERT that achieved AUC 70.31, MRR 34.89, nDCG@5 38.32, nDCG@10 43.95 [10]. Tao Qi and et al. in (2022) used News Recommendation with Candidate-aware User Modeling (CAUM) which achieved AUC 70.04, MRR 34.71, nDCG@5 37.89, nDCG@10 43.57 [11]. However, based on the previous achieved results, in our model we adopt GloVe algorithm for word embeddings and representation. Besides, the Multi-head Attention Layer calculates the attention of words, to generate a list of recommended news. We have compared our proposed systems vs studies and illiustaret the comparisions measurements in Table 2 in Result and discussion section.

## 3. Proposed System

In this section, we present our proposed system flowchart for the news recommendation system which consists of five phases, which we trained on a Large-Scale English Dataset for News Articles. Also, we discussed all the fifth phases and explained our contribution steps to the modified system (in the figure below). Our proposed system phases consist of:

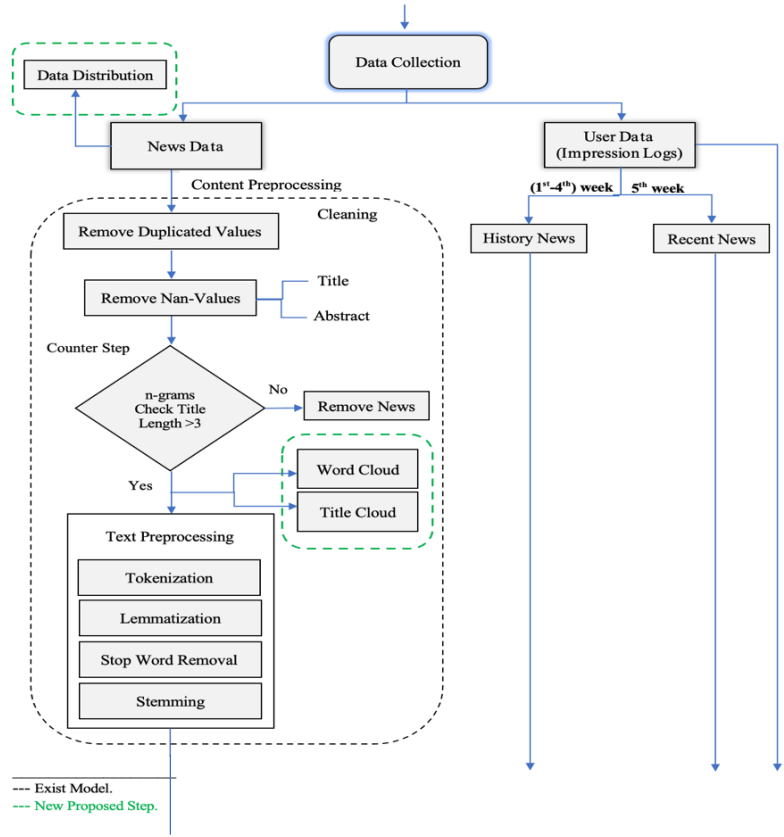

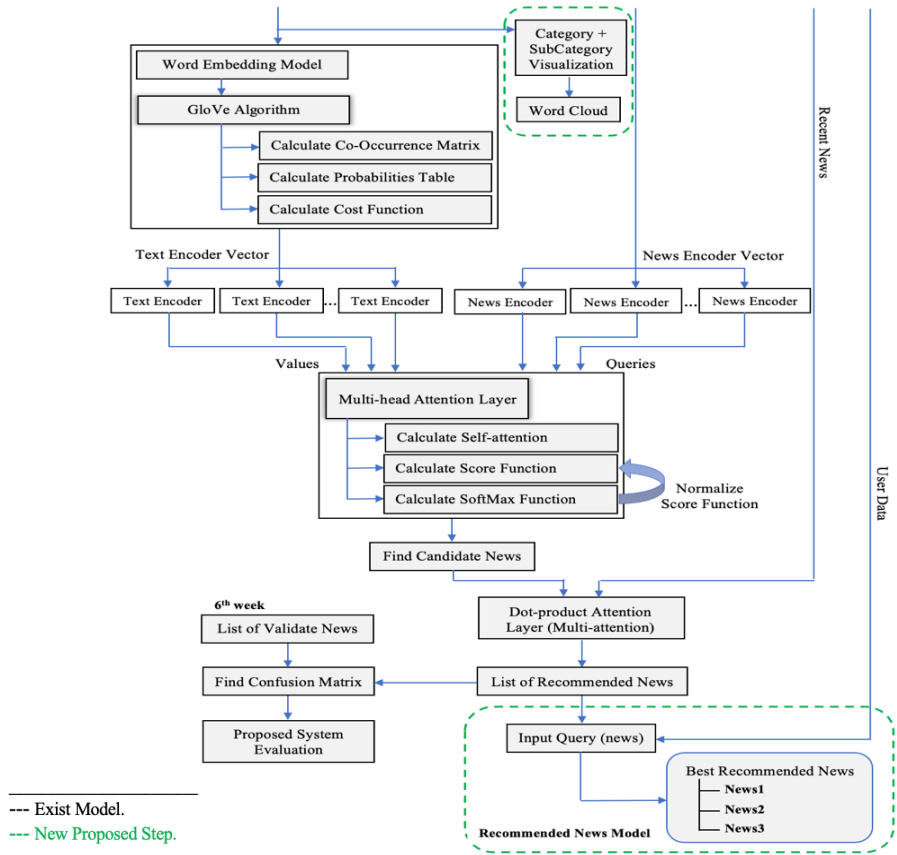

**Figure 1: The Proposed System Flowchart.**

### 3.1 Data Collection and Splitting Phase

This phase of the system is Data collection and splitting, which was collecting the news data from the Microsoft News platform for 6 weeks, and then splitting the data that was collected into News Data Corpus and User Data Corpus. Also, the User Data Corpus is separated into History News and Recent News.

### 3.2 Content Data Preprocessing Phase

This phase is Content Data Preprocessing, we have firstly used a cleaned algorithm for the content data and have three steps; remove the duplicated value, remove the nan-value of title or abstract, and the condition of checking the title length if less than or equal to 3 then remove the news else go to the next step. Then we used a preprocessing algorithm for the cleaned text that was produced from the first algorithm such as tokenization, lemmatization, stop word removal, and stemming.

### 3.3 Word Embedding and Representation Model Phase

This phase is Word Embedding and Representation Model by the **GloVe**[5] [12] algorithm which has three steps; calculate the co-occurrence matrix, calculate the probabilities table, and calculate the cost function. Then we encode the word embedding vector to the text encoder vector from Data News Corpus and encode the History News of the User Data Corpus.

---

[5] http://nlp.stanford.edu/projects/glove/.

GloVe equation is: $$J = \sum_{i,j=1}^{V} f(X_{ij})\left(w_i^T \widetilde{w}_j + b_i + \tilde{b}_j - \log X_{ij}\right)^2$$

### 3.4 Find Candidate News Phase

This phase is to Find the candidate news by **Multi-head Attention Layer** [13] which has three steps; calculate the self-attention, calculate the score function, and calculate the SoftMax Function. Then normalize the values of the score function based on these normalized values we can exclude irrelevant words. Finally, the results vector concatenates the words that are irrelevant to each other based on the attention value, and we get the candidate news.

SoftMax equation is: $$p_i = \frac{\exp(\hat{y}_i^+)}{\exp(\hat{y}_i^+) + \sum_{j=1}^{K} \exp(\hat{y}_{i,j}^-)}$$

### 3.5 Find List of Recommended News Phase

This phase is to Find a list of recommended news by the dot-product attention layer which multiplicated the values of candidate news and recent news, and calculates the alignment score function to find the correlation values between source and target. From this step, we get the list of recommended news.

### 4. Results Discussion

In this section, we evaluated our system and the results obtained using our proposed system in RS using the **GloVe** algorithm and **Multi-head Attention layer** against other systems of the news recommendation system in the same dataset version and showed the results with numbers and the year of the research. As shown in table 2:

**Table 2: Performance Comparison of Our Proposed System Against Other Systems Using MIND-Large Dataset Results.**

| No. | Study | Year of study | Model | Results | | | |
|---|---|---|---|---|---|---|---|
| | | | | AUC | MRR | nDCG@5 | nDCG@10 |
| 1 | U Kang [5] | 2020 | two NNG + four NNB | 0.7114 | 0.3568 | 0.3916 | 0.4485 |
| 2 | Danyang Liu and et al. [14] | 2020 | KRED | 0.6914 | | | 0.2684 |
| 3 | Shaina Raza and Chen Ding [15] | 2021 | D2NN | 0.538 | | | |
| 4 | Chuhan Wu and et al. [16] | 2021 | DA-Transformer | 68.32 | 33.36 | 36.34 | 42.07 |
| 5 | Chuhan Wu and et al. [17] | 2021 | UniRec | 68.41 | 33.50 | 36.47 | 42.26 |
| 6 | Chuhan Wu and et al. [8] | 2021 | PLMs | 70.64 | 35.39 | 38.71 | 44.38 |
| 7 | Yu Tian and et al. [18] | 2021 | KOPRA | 68.80 | 34.64 | 41.59 | 44.89 |
| 8 | Qi Zhang and et al. [7] | 2021 | UNBERT | 0.7068 | 0.3568 | 0.3913 | 0.4478 |
| 9 | Chuhan Wu and et al. [6] | 2021 | FedKD | 71.0 | 35.6 | 38.9 | 44.8 |
| 10 | Chuhan Wu and et al. [10] | 2021 | News-BERT | 70.31 | 34.89 | 38.32 | 43.95 |
| 11 | Chuhan Wu and et al. [19] | 2021 | Fastformer | 69.11 | 34.25 | 37.26 | 43.38 |
| 12 | Chuhan Wu and et al. [20] | 2021 | UaG | 69.23 | 34.14 | 37.21 | 43.04 |
| 13 | Jiahao Xun and et al. [21] | 2021 | IMRec | 0.6912 | 0.3364 | 0.3725 | 0.4364 |
| 14 | Peitian Zhang and et al. [22] | 2021 | SFI | 69.95 | 35.03 | 38.31 | 43.97 |
| 15 | Chuhan Wu and et al. [2] | 2021 | Fastformer+PLM-NR | 72.68 | 37.45 | 46.84 | 41.51 |
| 16 | Shuqi Lu and et al. [9] | 2021 | SEED-Encoder | 0.7059 | 0.3506 | 0.3908 | 0.4526 |
| 17 | Yu Song and et al. [3] | 2021 | pHUCB | 0.723 | | | |
| 18 | Chuhan Wu and et al. [23] | 2021 | LSTUR (random) | 68.76 | 33.94 | 36.89 | 42.55 |
| 19 | HAO SHI and et al. [24] | 2022 | DCAN | 0.5965 | | | 0.3243 |
| 20 | Tao Qi and et al. [25] | 2022 | ProFairRec | 67.64 | 33.08 | | 41.67 |
| 21 | Jian Li and et al. [4] | 2022 | MINER | 71.51 | 36.18 | 39.72 | 45.34 |
| 22 | Tao Qi and et al. [26] | 2022 | FUM | 70.01 | 34.51 | 37.68 | 43.38 |
| 23 | Tao Qi and et al. [11] | 2022 | CAUM | 70.04 | 34.71 | 37.89 | 43.57 |
| 24 | Rongyao Wang and Wenpeng Lu [27] | 2022 | MINS | 0.6811 | 0.3249 | 0.3601 | 0.4242 |
| 25 | Rongyao Wang and et al. [28] | 2022 | ANRS | 0.6826 | 0.3350 | 0.3722 | 0.4343 |
| **26** | **Our Proposed System** | **2022** | **GloVe+NRMS** | **71.211** | **35.72** | **38.05** | **44.45** |

Also, we have some new proposed steps in this system, which is Data Distribution to find the distribution of category and subcategory based on news count and will give an indicator of what news has been visited by users, and visualize the news that was collected in a graph.

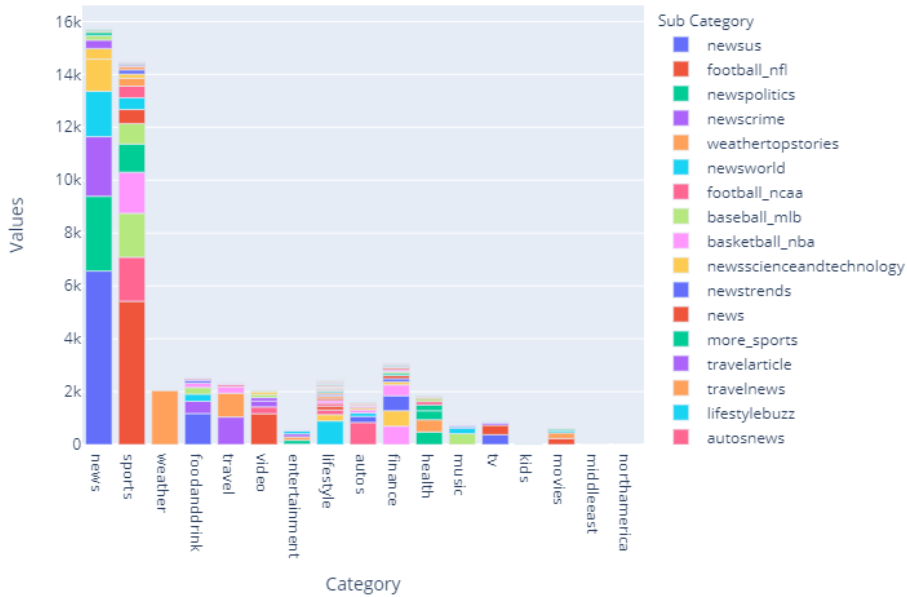

**Figure 2: Data distribution (Category, Subcategory).**

A new proposed step is Word Cloud to find the importance of words for each category. Also, Title Cloud to find the histogram of words for title length, and visualize it in a graph.

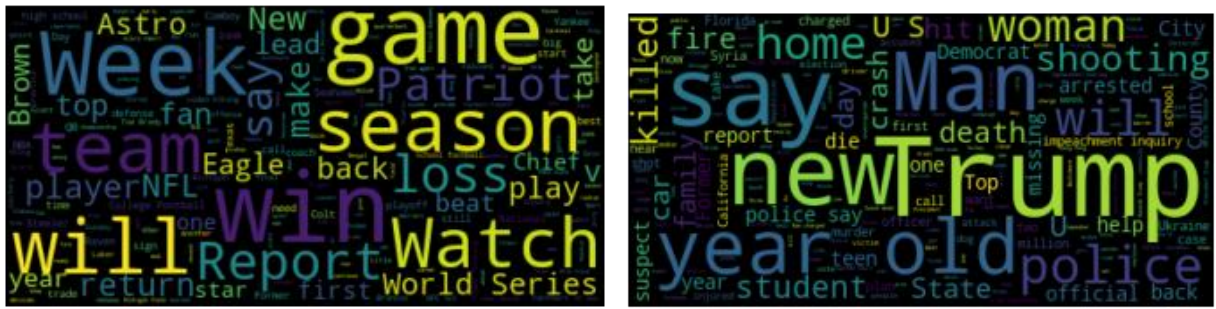

**Figure 3: (a) Word cloud (sports category).**      **(b) Word cloud (news category).**

And we find the title cloud which means the histogram of the words for title length, as the following figure:

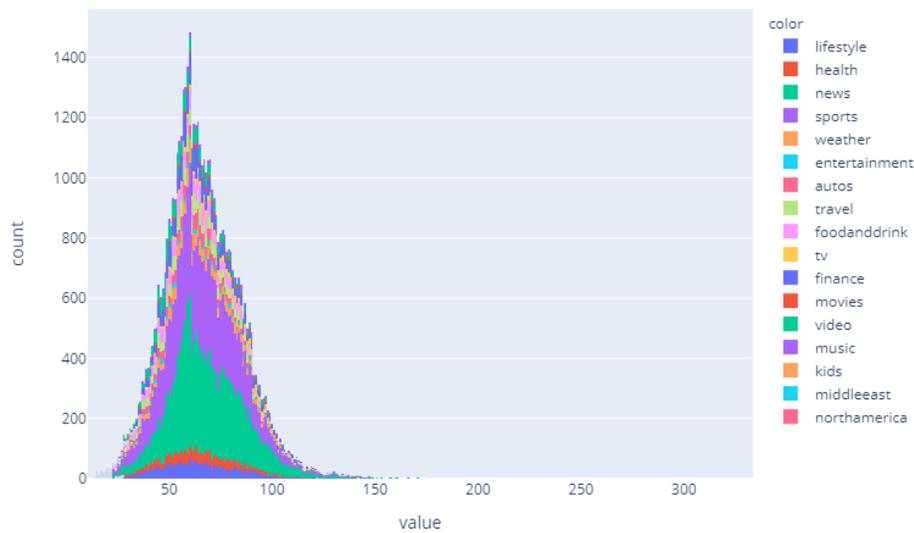

**Figure 4: Histogram for title length.**

Finally, a new proposed step is Recommended News Model for news recommendation retrieval and testing our model, which is based on pair-wise similarity distance between news vectors with the lowest value, and finding a list of recommended news.

```
News Title For Recommendation :How to Get Rid of Skin Tags, According to a Dermatologist
How to Get Rid of Skin Tags, According to a Dermatologist
=========================== News Article Name ===============================
News Headline :   How Get Rid Skin Tags , According Dermatologist

=========================== Recommended News :   ===============================
                                  headline Category                                           Abstract  similarity with the queried article
1                    Concerns about flat feet   health  Question: My child has flat feet. Should I be ...                            3.464102
2        Viagra Could Help Combat Blood Cancer Soon   health  Viagra might assist in making bone marrow tran...                            3.464102
3           Avoiding a Second Breast Cancer Surgery   health  Lumpectomies aren't perfect: Around 20% of the...                            3.464102
4             Headache Locations and their Meanings   health  Where it hurts may provide some clues as to wh...                            3.464102
5          Arthritis: Watch out for these symptoms   health  Characterized by inflammation in the joints or...                            3.464102
6      Alzheimer's can be prevented    Here's how...   health  We often read about how to reduce our chances ...                            3.464102
7                    Why your workout isn't working   health  For the average person who works out regularly...                            3.464102
8   President Carter out of surgery for subdural h...   health  Former President Jimmy Carter is recovering at...                            3.464102
9   What you should do if you break down on the Ca...   health  There are nearly 3,500 breakdowns a year on th...                            3.464102
10               Flu season is here in the Carolinas   health  Flu season is starting to take off in the Caro...                            3.464102
PS C:\NRMS>
```

**Figure 5: Results of our retrieval system.**

## 5. Conclusion

In this paper, we have a challenge with a new news recommendation system based on the MIND-Large version dataset. We proposed a modified system that consists of the GloVe technique of word embedding and two-layered of NRMS to learn the contextual word and news representations by modelling the interactions between words and news. We presented the implementation, results, and development steps of our proposed system of news recommendation system, which consists of five steps. And view the previous studies that are based on the same version of the dataset and compare them with our system. Finally, we get good results than some other related works.